\documentstyle[12pt]{article}

\newcommand{\vs}[1]{\rule[- #1 mm]{0mm}{#1 mm}}
\newcommand{\eq}{\vs{2}\begin{equation}}
\newcommand{\en}{\\[2mm]\end{equation}}
\newcommand{\bea}{\begin{eqnarray}}
\newcommand{\ena}{\end{eqnarray}}
\newcommand{\NP}[1]{Nucl.\ Phys.\ {\bf #1}}

\newcommand{\PL}[1]{Phys.\ Lett.\ {\bf #1}}

\newcommand{\AN}[1]{Ann. Phys. {\bf #1}}

\newcommand{\PR}[1]{Phys.\ Rep.\ {\bf #1}}
\newcommand{\PRev}[1]{Phys.\ Rev.\ {\bf #1}}
\newcommand{\PRL}[1]{Phys.\ Rev.\ Lett.\ {\bf #1}}

\begin{document}
\setcounter{page}{0}
\renewcommand{\thefootnote}{\fnsymbol{footnote}}

\rightline{IPNO/TH 93-15}

\vskip 3 true cm

\begin{center}
{\Large {\bf Experimental Determination }}\\
{\Large {\bf of Chiral Symmetry Breaking Parameters}}

\indent

\indent

{\large M. Knecht}

\indent

{\it Division de Physique Th\'eorique
\footnote{Unit\'e de Recherche des Universit\'es Paris XI et Paris VI
associ\'ee au CNRS.}, Institut de Physique Nucliaire\\
F-91406 Orsay Cedex, France} \\

\end{center}

\vskip 4 true cm

{\centerline{\bf Abstract}}

\indent

    In the low energy domain, Chiral Perturbation Theory parametrizes the small
chiral symmetry breaking effects, produced by the quark masses $m_u$, $m_d$ and
$m_s$, in terms of order parameters of massless QCD. The latter can then, in
principle, be measured in high precision, low energy experiments. We discuss
several relevant processes and possible improvements at future high luminosity
tau/charm or K factories.

\vskip 35 true mm

{\it Talk given at the XXVIIIth Rencontres de Moriond on QCD and High Energy
Hadronic Interactions, Les Arcs (France), March 20-27 1993.}

\indent

\newpage

%-----------------------------------------------------------------

\renewcommand{\thefootnote}{\arabic{footnote}}
\setcounter{footnote}{0}

The $SU(3)_L \otimes SU(3)_R$ chiral symmetry of QCD with three massless
flavours is spontaneously broken, producing an octet of massless pseudoscalar
Goldstone bosons. The observed masses $M_P$ of the light pseudoscalars then
result from the explicit breaking of chiral symmetry through the quark masses
$m_u$, $m_d$, $m_s$. The latter have values that are sufficiently small as
compared to the scale ${\Lambda}_H \sim 1$ GeV, where typical hadronic bound
states appear, so that their effects can be treated as perturbations around the
massless limit $m_u = m_d = m_s = 0$.

\indent

The masses of the pseudoscalars, for instance, have an expansion
\eq
{M_P}^{2} = B_{\bf 0} (m_i + m_j) + A_{\bf 0} {(m_i + m_j)}^2 + ...\ ,
\en
where the dots stand for higher order terms and non-analytic terms. The
constants $B_{\bf 0}$ and $A_{\bf 0}$ are not constrained by chiral symmetry,
but their
actual values convey informations about the ground state of QCD in the chiral
limit. As is well known, $B_{\bf 0}$ is related to the bilinear quark
condensate,
\eq
B_{\bf 0} = - {1\over{F_{\bf 0}^2}} <0\vert{\bar u}u\vert 0>
    = - {1\over{F_{\bf 0}^2}} <0\vert{\bar d}d\vert 0>
    = - {1\over{F_{\bf 0}^2}} <0\vert{\bar s}s\vert 0>\ ,
\en
where $F_{\bf 0}$ and $\vert 0>$ denote the pion decay constant and the vacuum
state, respectively, in the chiral limit. The constant $A_{\bf 0}$ also
corresponds to
an order parameter of chiral symmetry, namely to a two point function of scalar
and pseudo-scalar quark densities at vanishing momentum transfer, with the
infrared singularities arising from one and two Goldstone boson intermediate
states subtracted,
\eq
{\delta}^{ab} A_{\bf 0} = {{2i}\over{F_{\bf 0}^2}}\int dx <0\vert T\{ {\bar
q}_L
{{\lambda}^a\over{2}} q_R (x) {\bar q}_L
{{\lambda}^b\over{2}} q_R (0) + {\bar q}_R
{{\lambda}^a\over{2}} q_L (x) {\bar q}_R
{{\lambda}^b\over{2}} q_L (0)\}\vert 0 >^{(sub.)}\ .
\en
This two point function satisfies a superconvergent dispersion relation, whose
saturation with non-Goldstone hadronic states leads to an estimate
$A_{\bf 0}\sim
1-5$. Since colour degrees of freedom are confined, an estimation of the value
of $B_{\bf 0}$ along the same lines is not possible, while a direct
computation of
$B_{\bf 0}$ from QCD has not been achieved so far. $B_{\bf 0}$ could be of the
same order
as $\Lambda_H$ \cite{GOR}, or as small as $F_{\bf 0} \sim 90$ MeV \cite{FSS2}.
 Both scales appear as
natural, and our present theoretical understanding of QCD is consistent with
both possibilities. However, the two alternatives can, in principle, be
distinguished experimentally, as we shall discuss later.

\indent

For $B_{\bf 0}\sim\Lambda_H$, the first contribution dominates in Eq.(1), i.e.
the ratio
\eq
\eta\equiv {{(m_i + m_j )B_{\bf 0}}\over{{M_P}^2}}
\en
is of the order of unity, and the subsequent terms in Eq.(1) only give small
corrections to this value \cite{GL1}. For the same reason, the quark mass
ratio $r = 2m_s / (m_u + m_d )$ is close to its leading order value,
\eq
r \sim 2{{M_{K}^2} \over{M_{\pi}^2}} - 1 \sim 25\ .
\en
In the second alternative where $B_{\bf 0}\sim F_{\bf 0}$, the two terms in Eq.
(1) are comparable, unless the quark masses would be much smaller than usually
believed \cite{GL2}: $\eta$ may differ appreciably from
unity for $m_u$, $m_d$ in the range $10-50$ MeV, while the quark mass ratio
$r$ is no longer fixed by the lowest order mass formula, Eq.(1), and can be
considerably smaller than 25. An analysis \cite{FSS1} of the Dashen-Weinstein
sum rule for
the deviations from the Goldberger-Treiman relation shows that, unless the
pion-nucleon coupling constant were to differ by several standard
deviations \cite{Arn}
from the Koch-Pietarinen value \cite{Koch} $g_{\pi NN} = 13.40 \pm 0.08$, the
data indeed seem to
favour a value of $r$ in the range $10-15$. If confirmed, such a small value of
$r$ would mean that the standard
Chiral Perturbation theory converges very slowly.

\indent

In order to dispense with any a-priori idea concerning the value of $B_{\bf 0}$
, a mathematically consistent
rearrangement of the Chiral Perturbation Theory effective Lagrangian has been
proposed \cite{FSS2}, \cite{FSS3}, in
which $r$ (or $\eta$) appears as a free parameter. Roughly speaking, it amounts
to count a quark mass insertion as one power of external pseudoscalar momenta,
 whereas
in the standard case \cite{GL1}, the quark masses are counted like two powers
of momenta. For $r$ very different from
 25, one would then obtain a better convergence rate, while
for $r\sim$ 25, this improved Chiral Perturbation Theory reduces to the
standard one. It becomes then possible to show that the low-energy pseudoscalar
meson scattering amplitude is rather sensitive to the value of $r$. For
instance, the
leading order $\pi - \pi$ scattering amplitude reads \cite{FSS2}
\eq
A(s\vert t, u) = {{M_{\pi}}^2 \over{3{F_{\pi}^2}}}\alpha_{\pi\pi} +
                 {{s-{4\over 3}{M_{\pi}^2}}\over{{F_{\pi}}^2}}\beta_{\pi\pi}
\en
where
\eq
\beta_{\pi\pi} = 1\ ,\ \ \alpha_{\pi\pi} = 1 + 6\, {{r_2 - r}\over{r^2 - 1}}\ ,
\en
and $r_2$ denotes the quantity appearing in Eq.(5). In the standard case
$B_{\bf 0}
\sim\Lambda_H$, $r = r_2$ at leading order and one recovers Weinberg's theorem
\cite{W}.
The extreme case $B_{\bf 0} = 0$ gives $\alpha_{\pi\pi} = 4$, while for
$r\sim 10$,
one obtains $\alpha_{\pi\pi} \sim 2$. Hence, the value of $r$ can, in
principle, be
obtained from $\pi -\pi$ scattering data. The authors of Ref. \cite{FSS3} have
considered the possibility to extract the value of ${{\alpha_{\pi\pi}} /
\beta_{\pi\pi} }$ from existing data. The lack of {\it sufficiently precise}
experimental information concerning, for instance, the I=0 s-wave scattering
length, $a_0^0=0.26\pm 0.05$ (see e.g. Ref. \cite{N} ),
allows the ratio ${{\alpha_{\pi\pi}} /
\beta_{\pi\pi} }$ to vary within a rather large range $\sim 1.5 - 4$, which
could accomodate any value of $B_{\bf 0}$ in the range between $F_{\bf 0}$ and
1 GeV.
To settle the fundamental question of the value of $B_{\bf 0}$, more precise
data on
the $\pi - \pi$ phase shifts are needed. At the same time, one should explore
other experimental situations where informations on $r$ could be obtained.

\indent

The above discussion reflects a rather general situation: no {\it accurate}
(at a precision level of a few percents) experimental informations on chiral
symmetry breaking effects are available at present. This is quite
understandable, since
usually they only represent an almost negligible fraction of the total
observed effect. In this respect, future high-luminosity K factories and/or
tau/charm factories offer some interesting opportunities to improve the
experimental situation.

The next generation of $K_{e4}$ experiments at DA$\Phi$NE will increase the
number of events by a factor of ten within one year of data taking, and thus
improve the precision on the $\pi - \pi$ phase
shifts (for a discussion of semi-leptonic K decays, see Ref. \cite{BEG} ). A
determination of $a_0^0$ within, say, 5\% would rule out the scenario
$B_{\bf 0}\sim \Lambda_H$ if the present central value of $0.26$ would be
confirmed.

An interesting possibility to pin down the value of $r$ directly in $K_{\mu
4}$ decays has also been suggested \cite{FKSS}. It relies on the observation
that the axial vector
form factor $R$ (in the notations of Ref. \cite{BEG} ), whose contributions to
the
amplitude appear with a suppression factor $m_e^2$ in the case of $K_{e4}$
decays, contributes substancially in the case of $K_{\mu 4}$ decays, and
exhihits an observable dependence with respect to $r$ at leading order.

The analysis of $D_{l4}$ decays (e.g $D^+ \to K^-\, \pi^+\, e^+\, \nu_e$) at a
tau/charm factory could similarly lead to precise determinations of the phase
shifts for $K-\pi$ scattering, which has also been discussed within Chiral
Perturbation Theory \cite{B}, \cite{BKM}.

Quite generally, a tau/charm factory would be welcome as a source of light
flavours, through the hadronic decays $\tau \to n\pi\ \nu_{\tau}$, or $\tau\to
n\pi\  K\ \nu_{\tau}$. The extraction of the chiral symmetry breaking scalar
form factor in e.g. $\tau\to 3\pi\  \nu_{\tau}$ is kinematically possible
\cite{KM}, even with non polarized $\tau$'s, but whether it can be done in
practice with sufficient accuracy remains unclear. It is also interesting to
note that a precise normalization of the divergence of the axial vector in this
process would allow for a determination of the quark masses themselves, through
the evaluation of the QCD spectral sum rules \cite{DeR}.

Finally, the possibility to obtain the value of the combination $a_0^0 - a_0^2$
of $\pi - \pi$ scattering lengths from the measure of the lifetime of $\pi^+ -
\pi^-$ atoms \cite{Nem} should also be mentioned. A letter of intent \cite{Cz}
for such an experiment at CERN has recently been approved \cite{Mon}.

\indent

\noindent{\bf Acknowledgements}

I thank Jan Stern for several illuminating discussions and for a careful
reading of the manuscript, Lucien Montanet for an informative conversation
concerning the ${\pi}^+ - {\pi}^-$ atom project at CERN, and the Organizers
of the ``Rencontres" for creating a stimulating atmosphere.

\indent

\end{document}